\documentclass[twocolumn]{aastex3}

\usepackage{latexsym, graphicx, amssymb, longtable, epsf, natbib}
\usepackage{amssymb}
\usepackage{amsmath}
\usepackage{epstopdf} 

\usepackage{paralist}

\DeclareMathOperator\arctanh{arctanh}

\newcommand{\Alfven}{Alfv\'{e}n }
\newcommand{\Alfvenic}{Alfv\'{e}nic }
\newcommand{\PSP}{Parker Solar Probe }
\newcommand{\SO}{Solar Orbiter}

\newenvironment{innerlist}[1][\enskip\textbullet]%
        {\begin{compactenum}[#1]}{\end{compactenum}}

\slugcomment{Astrophysical Journal, Accepted 1 August 2017}

\begin{document}


\title{A zone of preferential ion heating extends tens of solar radii from Sun}



\author{J. C. Kasper\altaffilmark{1,2}, K.~G. Klein\altaffilmark{3}}
\affil{University of Michigan, Michigan, USA}

\author{T. Weber}
\affil{U Boulder, Boulder, USA}

\author{M. Maksimovic, A. Zaslavsky}
\affil{Laboratoire d'Études Spatiales et d'Instrumentation en Astrophysique, Observatoire de Paris-CNRS-Université Pierre et Marie Curie-Université Denis Diderot, Meudon, France}


\author{S. D. Bale}
\affil{Space Sciences Laboratory and Physics Department, University of California, Berkeley, California, USA}

\author{B. A. Maruca}
\affil{Department of Physics and Astronomy, University of Delaware, Delaware, USA}

\author{M. L. Stevens, A. W. Case}
\affil{Harvard-Smithsonian Center for Astrophysics, Cambridge, USA}


\altaffiltext{1}{jckasper@umich.edu}
\altaffiltext{2}{Smithsonian Astrophysical Observatory}
\altaffiltext{3}{Lunar and Planetary Laboratory, University of Arizona}



\begin{abstract}
The extreme temperatures and non-thermal nature of the solar corona
and solar wind arise from an unidentified physical mechanism that
preferentially heats certain ion species relative to others.
Spectroscopic indicators of unequal temperatures commence within a
fraction of a solar radius above the surface of the Sun, but the outer
reach of this mechanism has yet to be determined.  Here we present an
empirical procedure for combining interplanetary solar wind
measurements and a modeled energy equation including Coulomb
relaxation to solve for the typical outer boundary of this zone of
preferential heating.  Applied to two decades of observations by the
Wind spacecraft, our results are consistent with preferential heating
being active in a zone extending from the transition region in the
lower corona to an outer boundary 20-40 solar radii from the Sun,
producing a steady state super-mass-proportional $\alpha$-to-proton
temperature ratio of $5.2-5.3$.  Preferential ion heating continues
far beyond the transition region and is important for the evolution of
both the outer corona and the solar wind.  The outer boundary of this
zone is well below the orbits of spacecraft at 1 AU and even
closer missions such as Helios and MESSENGER, meaning it is likely
that no existing mission has directly observed intense preferential heating,
just residual signatures.  We predict that {Parker Solar Probe} will
be the first spacecraft with a perihelia sufficiently close to the Sun
to pass through the outer boundary, enter the zone of preferential
heating, and directly observe the physical mechanism in action.
\end{abstract}




\section{Introduction} \label{sec:intro}

Observations of space over the last half century, including
spectroscopic diagnostics of UV emission from coronal plasma and
direct in situ sampling of solar wind by spacecraft have shed light on
the non-thermal nature of heating in the corona and solar
wind. Throughout the heliosphere, plasma is typically found in states
other than local thermodynamic equilibrium, with relative drifts and
unequal temperatures between species and anisotropic and otherwise
non-Maxwellian velocity distribution functions commonly observed. Such
non-thermal structure is indicative of mechanisms that selectively couple to particles with particular velocities, charges or masses and preferentially heat
different plasma species. One region in particular where our
understanding of these mechanisms is incomplete is the inner
heliosphere.

The visible $6000\ K$ photosphere of the Sun is surrounded by a $1-10
\ MK$ solar corona that reaches many solar radii ($R_s$) into space
before transitioning into the supersonic and ultimately
super-\Alfvenic solar wind. The temperature of the solar atmosphere
rises to $10^5 \ K$ within several hundred kilometers in the narrow
transition region at the base of the corona. At around $0.1-0.3 R_s$,
rising temperatures $T$ and falling densities $n$ are such that the
frequency of Coulomb collisions, $\nu_c\propto n/T^{3/2}$, drops to
the point that the coronal plasma becomes effectively collisionless,
with electrons and individual ion species not persisting in a common
local thermodynamic equilibrium. Ions become much hotter than
electrons, and heavier ions achieve higher temperatures than the
hydrogen that composes the majority of the coronal plasma
\citep{Esser:1999,Landi:2009}.  Emission has been detected from steady
non-flare coronal oxygen at $\sim 10^8 K$, a hundred times hotter than
coronal hydrogen and more than six times hotter than the core of the
Sun. Such unequal temperatures serve as a signature of preferential
heating of different species in the corona. It is possible that
preferential heating is occurring lower in the solar atmosphere, but
the higher frequency of Coulomb collisions at lower heights would
remove the signature of such heating. The relative temperatures of ion
species are highly variable, with a statistical preference for either
equal temperatures or equal thermal speeds corresponding to
mass-proportional temperatures, but intermediate temperatures and
super-mass-proportional temperatures are also observed. For example, a
recent study suggested that coronal ions develop an equilibrium
temperature $T_i/T_p\approx (4/3) m_i/m_p$ \citep{Tracy:2016}.  One of
the most significant open challenges in solar and space physics is to
unambiguously determine the physical processes responsible for this
heating.

There are many plausible theories for the physical processes
responsible for the extended and preferential heating of different ion
species in the corona and solar wind, including resonant wave heating
\citep{Cranmer:2000,Hollweg:2002}, velocity filtration
\citep{Scudder:1992a}, impulsive events including reconnection
\citep{Cargill:2004,Drake:2009}, and stochastic heating by
low-frequency \Alfvenic turbulence
\citep{Chandran:2010a,Chandran:2010b}.  Unambiguous
  identification of the dominant process is complicated by uncertainty
  in the nature of energy readily available for dissipation in the
  corona.  For example, high frequency waves could escape from the
  photosphere through the transition region before being damped in the
  lower corona \citep{Axford:1997}.  Alternately, MHD
  turbulence could be generated locally by the interaction between
  outward and reflected low frequency waves anywhere below the solar
  wind Alfv\'en point \citep{Matthaeus:1999b}.  Recent work
\citep{Kasper:2007,Chandran:2013,Kasper:2013} has shown that velocity
moments of solar wind H$^+$ and He$^{2+}$ ions are consistent with both strong
heating due to resonant absorption of Alfv\'en-cyclotron waves or
stochastic heating. This heating could persist throughout the
heliopshere or could occur only in a select region near the Sun, with
the resultant non-thermal structure being reduced by infrequent
Coulomb collisions as the solar wind expands
\citep{Kasper:2008,Tracy:2016}.  In situ measurements over the last
half-century, including those from the twin {Helios} spacecraft that
approached to within $62 R_{\rm S}$ from the Sun, show that the radial
gradients of ion and electron temperatures are much more shallow than
would be expected from a cooling solar wind undergoing adiabatic
expansion \citep{Hellinger:2013}. This evidence for ongoing radial
heating in the inner heliosphere is not necessarily evidence for
ongoing preferential ion heating of the type observed near the
Sun. More detailed tests involving the correlation of particle
distribution function structure and electromagnetic fields may be able
to identify the energy source and distinguish between
the proposed mechanisms \citep{Klein:2016a}, but such tests need
measurements of the plasma as the heating is occurring.  It is
therefore important to determine how far away from the Sun the
preferential heating mechanism is active, and thus how close to the
Sun a spacecraft must approach to directly resolve the process.

In this paper, we address three related questions: Are unequal
temperatures in the solar wind maintained by ongoing local
preferential heating, or are they a leftover of heating that happened
close to the Sun?  Is faster solar wind further from local
thermodynamic equilibrium than slow wind because only fast wind
experiences preferential heating in the corona resulting in
non-thermal structure? How far from the Sun does preferential ion
heating continue?  The purpose of this paper is to develop a technique
for measuring how much time has elapsed since solar wind ions
experienced preferential heating that was sufficiently strong to
generate super-mass-proportional temperatures.  We assume that there
is a zone of preferential heating surrounding the Sun, starting at
$0.1-0.3 \ R_s$ as indicated by spectroscopic observations, and ending
at an outer boundary $R_b$. Beyond $R_b$ any remaining
non-preferential heating is weak and Coulomb collision dominates,
leading to an exponential decay of the temperature ratio $T_i/T_p$
toward unity. Observational motivation for this work is presented in
Section~\ref{sec:obs}, with a model for the radial evolution of the
temperature ratio between H$^+$ and He$^{2+}$ detailed in
Section~\ref{sec:model}. The technique for measuring the outer
boundary of the zone combining the derived model and in situ
measurements from the {Wind} spacecraft is presented in
Section~\ref{sec:Rb}. Values for the outer boundary,
Section~\ref{sec:discuss}, are found to be within the perihelion of
Parker Solar Probe, allowing for verification or falsification of our model and,
potentially, the first in situ observation of the preferential heating mechanism.

\section{Observations of Coulomb Thermalization} \label{sec:obs}

The observational basis of this work is an extensive set of
measurements of solar wind plasma collected by the Solar Wind
Experiment (SWE, \citet{Ogilvie:1995}) and the Magnetic Field
Investigation (MFI, \citet{Lepping:1995}) instruments on the NASA
{Wind} spacecraft. {Wind} was launched in late 1994 and has operated
continuously in a variety of orbits passing through the solar wind
near Earth, resulting in a comprehensive set of observations of solar
wind in the ecliptic plane spanning nearly two decades.  Solar wind
H$^+$ (protons) and He$^{2+} \ (\alpha$ particles) are measured by the
SWE Faraday Cup instruments, which record a detailed three-dimensional
measurement of the velocity distribution function (VDF) of the two ion
species once every 90 seconds.  We use a technique developed to
extract anisotropic temperatures and differential flows for each
species as first described by \citet{Kasper:2002}. This algorithm
makes use of 3-second time resolution measurements of the vector
magnetic field by the MFI flux gate magnetometers in order to
determine the temperature of each ion species parallel and
perpendicular to the local magnetic field. Approximate uncertainties
in the resulting observations were documented in \citet{Kasper:2006},
which estimated a typical uncertainty in an ion temperature
measurement of 8\%. We follow the same data selection procedures
described in \citet{Kasper:2008}, but with an additional 8 years of
observations.

Previous work \citep{Feldman:1974,neugebauer76,livi86,Kasper:2008,Tracy:2016} has
demonstrated that Coulomb relaxation plays an important role in
thermalizing the solar wind ions, removing non-thermal structure such
as temperature anisotropy and temperature disequilibrium between
species. The effect of this thermalization can be quantified by the
estimated number of Coulomb thermalization times that have elapsed in
the time it takes for the solar wind to expand from the corona to the
observing spacecraft, a quantity often referred to as the Coulomb
collisional age $A_c$ \citep{Salem:2003,Kasper:2008,Maruca:2013b}. We
will reserve $A_c$ for a more precise calculation presented later in
this paper, and introduce the Coulomb number $N_C=\nu_{ab} R/U$ to
indicate a rough approximation based only on observations at a
spacecraft in interplanetary space with no accounting for propagation
effects.  In this equation $R$ is the total distance of the spacecraft
from the Sun, $U$ is the speed of the solar wind, and $\nu_{ab}$ is
the characteristic rate for Coulomb interactions between two species
$a$ and $b$; for $N_C$ both $\nu_{ab}$ and $U$ are assumed to be
constant. Throughout this paper we make use of the calculations of
\cite{hernandez87} for the Coulomb interaction between two species
with Maxwellian distribution functions, different temperatures,
densities, and a net differential flow as discussed in more detail in
the following section.

\begin{figure*}
  \includegraphics[width=7in]{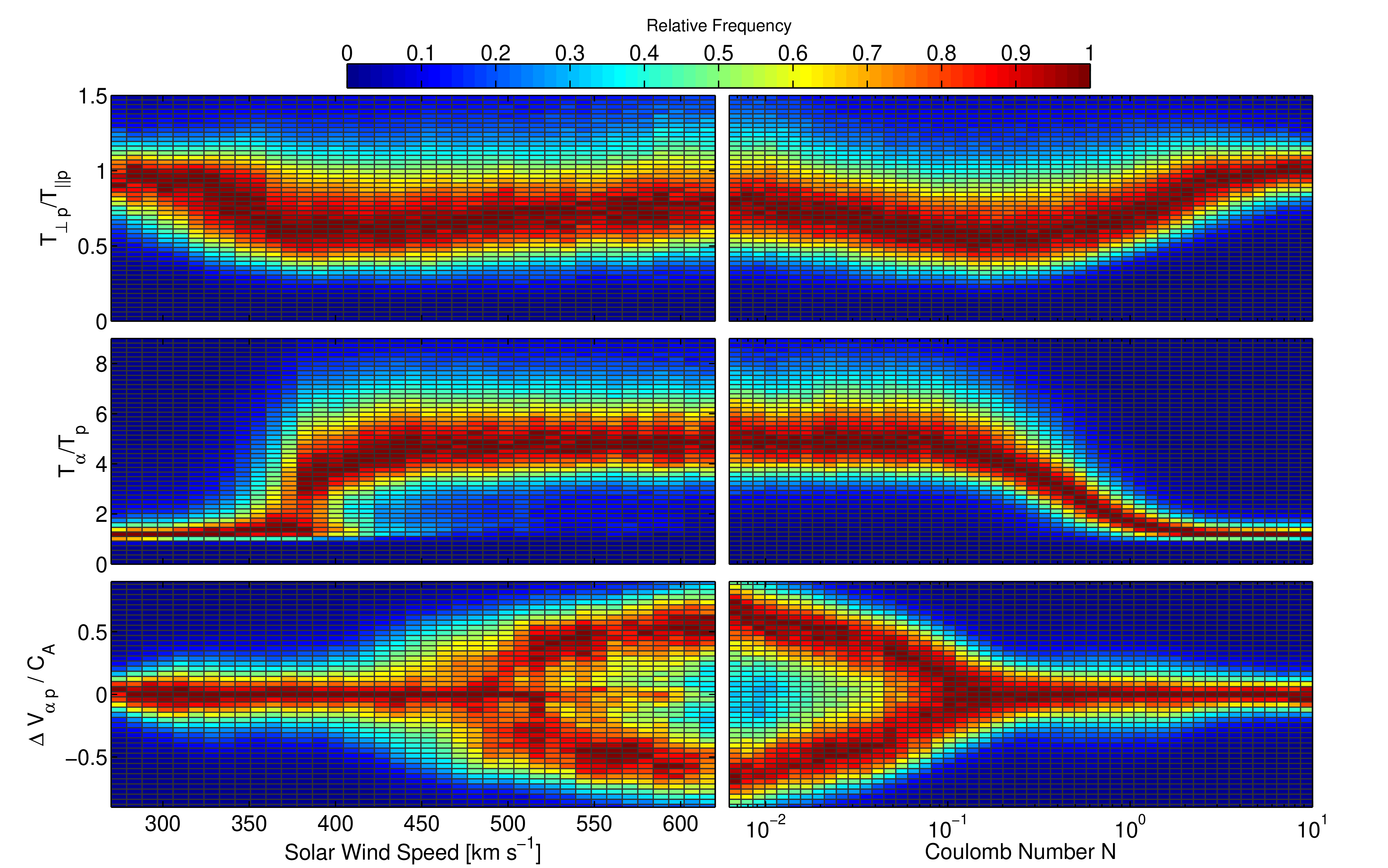}
  \caption{Two dimensional histograms of the distributions of
    $T_{\perp p}/T_{\parallel p}$, $T_\alpha/T_p$, and $\Delta
    V_{\alpha p}/C_A$ as functions of solar wind speed $U$ (left) and
    Coulomb number $N_C$ (right).  While non-thermal solar wind is
    generally associated with high speeds, these distributions suggest
    that the occurrence frequency is really determined by the Coulomb
    number $N_C$.}
\label{fig:ageorspeed}
\end{figure*}

Column normalized distributions of three markers of non-thermal
structure, $T_{\perp p}/T_{\parallel p}$, $T_\alpha/T_p$, and the
$\alpha$-proton drift velocity normalized by the \Alfven speed $\Delta
V_{\alpha p}/C_A$, are plotted as a function of solar wind speed $U$
in the left panels of Fig.~\ref{fig:ageorspeed}. As has been reported
many times before, the fast solar wind is more non-thermal than slow
solar wind (c.f. \citet{Marsch:2012}).  In the right three panels, the
same markers are plotted as a function of $N_C$.  As previously
reported in \cite{Kasper:2008}, $N_C$ is shown to also order the
observations of these three non-thermal measures. The magnitudes of
the properties are observed to decrease exponentially with both $U$
and $N_C$ for sufficiently large $N_c$. For instance,
the dependence of $T_\alpha/T_p$ on $N_c$ is monotonic, with a single
peak value of the temperature ratio for each value of $N_C$ which
decreases exponentially with large Coulomb number and a normal
distribution of temperature ratios about that peak. The same cannot
be said of the dependence of $T_\alpha/T_p$ on $U$; while they are
also strongly correlated, the spread in each $T_\alpha/T_p$ is larger
and has multiple peaks for a given value of $U$.  In
  general, the distribution of $T_\alpha/T_p$ for a given $U$ is
  further from a normal distribution than the distribution for a given
  $N_C$. The variation of $T_{\perp p}/T_{\parallel p}$ and $\Delta
  V_{\alpha p}/C_A$ are more complex, possibly because they are more
  sensitive to kinetic microinstablities and the effects of expansion,
  but even so these non-thermal features are washed away at
  sufficiently large $N_c$.
 
The exponential dependence of $T_\alpha/T_p$ on $N_C$ is consistent
with a simple model for the radial evolution of the temperature ratio.
Considering the thermalization of temperature differences in the
absence of any effects other than Coulomb collisions, keeping $T_p$
constant, and following \citet{Spitzer:1962}, the time evolution of
$T_\alpha/T_p$ can be written as $d (T_\alpha/T_p) /d t =
-\nu_{\alpha,p} T_\alpha/T_p$, yielding a solution of $T_\alpha/T_p
\sim \exp \left[ - \int \nu_{\alpha,p}dt \right]$.  Under the
oversimplifying but instructive assumption that $\nu_{\alpha p}$ is
constant, and that the appropriate dynamical time is the transit time
from the Sun, allows the further simplification $T_\alpha/T_p \sim
\exp \left[ - N_c \right]$. This form is in good agreement with the
solar wind observations, raising several interesting possibilities.
First, the fact that this single formula fits all of the Wind
observations across all solar wind speeds suggests that non-unity
$T_\alpha/T_p$ and preferential ion heating may not be restricted to
fast solar wind.  Perhaps all solar wind close to the Sun experiences
strong preferential ion heating and develops a large $T_\alpha/T_p$,
and the apparent association between $T_\alpha/T_p$ and $U$ is simply
due to the fact that the number of Coulomb collisions a parcel of
solar wind experiences varies strongly with $U$.  Slower wind both
takes longer to get to the spacecraft and tends to have a
significantly higher $\nu_{\alpha p}$, resulting in a stronger
suppression of non-thermal $T_\alpha/T_p$ which may be present closer
to the Sun. One might counter that $T_\alpha/T_p$ and $N_C$ are both
strongly correlated with speed or temperature, giving the false
impression that $N_C$ regulates $T_\alpha/T_p$.  However,
\cite{Maruca:2013b} demonstrated that the temperature ratio has a
stronger correlation with the number of Coulomb collisions than other
solar wind parameters such as density, speed, and temperature.

\begin{figure}
	\includegraphics[width=2.5in,angle=90]{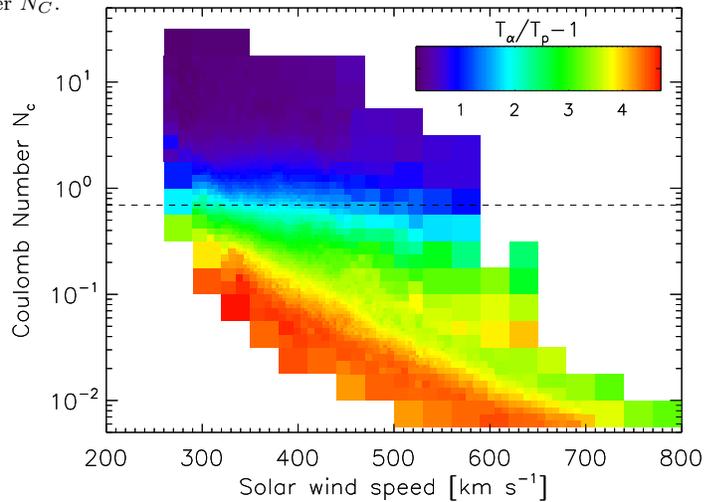}
  \caption{Excess temperature $\epsilon=T_\alpha/T_p-1$ as a function
    of the solar wind speed $U$ and Coulomb number $N_c$. For all
    solar wind speeds, $\epsilon$ decays exponentially with increasing
    $N_C$, falling to about half its maximum value near $N_C \sim 0.7$
    as one would expect from a simple relaxation process.}
\label{fig:tap_vp_ac}
\end{figure}

To further show that $N_c$ and $T_\alpha/T_p$ are not simply dependent
on $U$, we plot in Fig. \ref{fig:tap_vp_ac} the mean value of the
excess temperature ratio $\epsilon \equiv T_\alpha/T_p-1$ as a
function of both solar wind speed and $N_C$. While there is clearly a
trend in the typical $N_C$ as a function of speed, the exponential
drop in $\epsilon$ from high values of $\approx 4$ to less than unity
happens at all observable speeds at $N_C\sim 0.7$, with slight
dependence on speed.  Even the slowest solar wind, with speeds less
than $300 \ {\rm km \ s^{-1}}$, has high $\epsilon$ when $N_C$ is
small.  As captured in Figs.\ref{fig:ageorspeed} and
\ref{fig:tap_vp_ac}, observations of the solar wind are consistent
with a hypothesis that all plasma close to the Sun experiences
preferential heating of ions, even plasma that results in slow wind.
The association of significant $\epsilon$ with faster wind speeds is
simply due to the fact that slower wind in general has a higher
collisional age, leading to a removal of the non-thermal structure by
the time the plasma reaches 1 AU. This result is significant, because
it implies that mechanisms that could produce non-thermal heating may
be active both in slow and fast wind.  In the following sections we
use this theoretical framework to produce an estimate of how far from
the Sun this heating occurred.

\section{Modeling the Preferential Heating Zone} 
\label{sec:model}

The clear exponential dependence of $\epsilon$ on $N_C$ in
Figs. \ref{fig:ageorspeed} and \ref{fig:tap_vp_ac} is suggestive of
the gradual thermalization due to Coulomb relaxation on a non-thermal plasma.
\citet{Spitzer:1962} showed that non-thermal plasma relaxes to thermal
equilibrium through a series of small-angle scattering of ions
mediated by the Coulomb interaction.  In the absence of any other
processes, two species with a temperature difference $\Delta T$ will
come in to equilibrium at a rate $d \Delta T/dt=-\nu_c \Delta T$.
Ignoring any $T$ dependence in $\nu_c$, we can rearrange this equation
as $d\Delta T/\Delta T=-\nu_c dt$, or integrating both sides and
exponentiating,
\begin{equation}
\Delta T=\Delta T_o e^{-\int \nu_c dt}
\end{equation}
where we can define $\Delta T_o$ as the initial temperature difference
and the collisional age $A_c$ of the plasma as that integral over time
of all Coulomb collisions experienced by the plasma since it began to
relax
\begin{equation}
A_c\equiv\int \nu_c dt\simeq N_C.
\end{equation}

We now develop a more sophisticated model for the behavior of
$T_\alpha/T_p$, which improves upon the assumption that $\nu_{\alpha
  p}$ is constant and that the correct dynamical time is the transit
time from the center of the Sun to Earth at constant speed as used for
$N_c$. Such an approach was used in \citet{Maruca:2013b}, which solved
the ion temperature differential equations backwards in time to
investigate the distribution of $\epsilon$ near the Sun, finding that
for radial distances of $0.1 \ AU$ $\epsilon$ took on highly non-thermal
values for all solar wind speeds.  In this paper we do the opposite;
we assume that the plasma is highly non-thermal near the Sun, with a
large value of $\epsilon$ below some radial boundary $R_b$, and that
the observed variation in $\epsilon$ at 1 AU is subsequently
determined solely by Coulomb relaxation.  Values for $R_b$ are then
obtained from comparing models for radial solar wind behavior with in
situ measurements at $1 \ AU$.

\begin{figure}
  \centerline{\includegraphics[width=2.75in, angle=90]{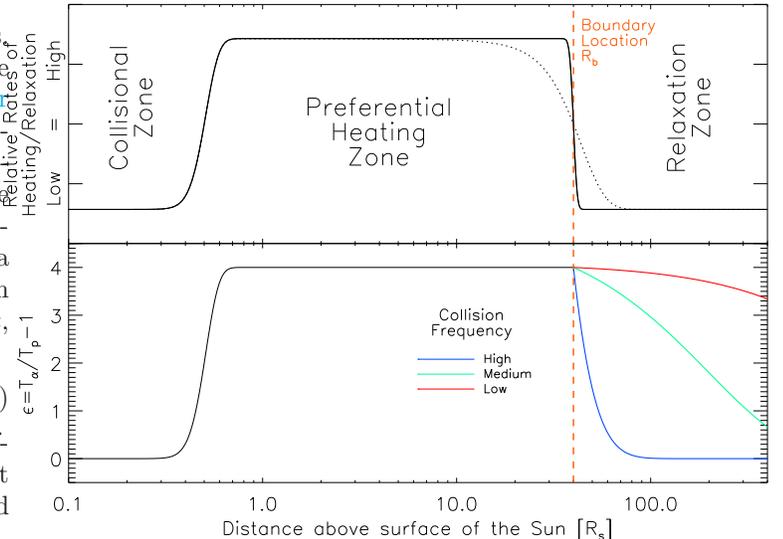}}
  \caption{Our simple three-zone model for ion temperature ratios in
    the inner heliosphere.  The upper panel schematically indicates
    the ratio of the relative rates of preferential ion heating to
    Coulomb relaxation as a function of distance.  The lower panel
    indicates the resulting excess temperature of He$^{2+}$ relative
    to $H^+$.  Close to the surface of the Sun the plasma is highly
    collisional and isothermal with $\epsilon=T_\alpha/T_p-1=0$.
    Above some height collisions are inefficient and $\epsilon$ rises
    to an equilibrium value. At the outer boundary $R_b$ the
    preferential heating stops, and $\epsilon$ decays exponentially
    with time proportional to the collision frequency.  The observed
    value of $\epsilon$ at a spacecraft such as Wind is then a
    function of the equilibrium value in the preferential heating zone
    and the effects of collisions integrated from $R_b$ to the
    observer.}
\label{fig:regions}
\end{figure}

We make the following key assumptions in the construction of our
model, which are are illustrated schematically in
Fig.~\ref{fig:regions}:
\begin{innerlist}
\item There is a zone in the inner heliosphere where the Coulomb
  collision frequency is sufficiently low and the ion heating rate,
  due to unspecified mechanisms, is sufficiently high to allow for
  preferential heating of ions.  Based on spectroscopic observations
  of ion temperatures in the corona this zone begins just $0.2-0.3 \
  R_s$ above the photosphere, but the outer extent of this zone is
  unknown.
\item The preferential heating results in different ion temperatures,
  with $\epsilon$ reaching an asymptotic value $\epsilon_0$ within the
  zone.  Here we are motivated by the fact that the observed spread in
  $\epsilon$ is very narrow for small $N_c$.
\item We assume that at some distance from the Sun the preferential
  heating falls off and quickly becomes negligible.  We define
  this outer boundary of the zone as $R_b$.  
\item Above $R_b$, $\epsilon$ decays exponentially as a function of
  the number of Coulomb collisions.
\end{innerlist} 
We acknowledge that this model makes several critical simplifications,
each of which merits further investigation. For example, $R_b$ may
vary with time, solar wind type, or level of solar activity.  The
preferential heating in practice will not shut off completely at
$R_b$, and it would be worthwhile to investigate the impact of a more
gradual evolution.  Finally, we know that the steady state $\epsilon$
in solar wind with low $A_c$ is a function of other plasma properties,
such as differential flow and plasma $\beta$ \citep{Kasper:2013}, and
has a non-negligible spread for a given set of
parameters. Nonetheless, for the purposes of this paper, where we aim
to determine if the mean value of the observed temperature excess can
be described using a fixed outer boundary, and differentiate between a
boundary in the lower corona, interplanetary space, or somewhere in
between, this model is sufficient.
 
To model the excess temperature, we start with an
energy equation for $T_p$ and $T_\alpha$
\begin{equation}
\frac{d T_s}{d r} = 
\left(\gamma - 1 \right)
\left[\frac{T_s}{n_s}\frac{d n_s}{d r}
-\frac{Q_s}{n_s k_B U} \right]
-\sum_{s'}\frac{\nu_{ss'}}{U}\left(T_{s}-T_{s'}\right),
\label{eqn:Ts}
\end{equation}
which includes the effects of expansion, input heating, and
collisional relaxation. The Coulomb coupling between the ion species
is governed by the frequency of energy-changing collisions between the
two species $\nu_{ss'}$, and the input heating is parameterized by the
heat input rate $Q_s$ in $\rm{ergs \ s^{-1} \ cm^{-3}}$. This form of
the adiabatic energy equation, found for example in
\cite{Cranmer:2009}, assumes a steady wind with radially dependent
speed of $U(r)$.  Given this form of radial temperature evolution, and
assuming the collisional coupling is dominantly between the protons
and $\alpha$ particles, the radial change in $\epsilon$ is
\begin{align}
\begin{split}
\frac{d \epsilon}{d r} =& \frac{1}{T_p}\frac{d T_\alpha}{d r} 
- \frac{T_\alpha}{T_p^2}\frac{d T_p}{d r} \\
= & \frac{\left(\gamma - 1 \right)}{T_p}
\left[\frac{T_\alpha}{n_\alpha}\frac{d n_\alpha}{d r}
-\frac{T_\alpha}{n_p}\frac{d n_p}{d r}
-\frac{Q_\alpha}{n_\alpha k_B U}
+\frac{T_\alpha}{T_p}\frac{Q_p}{n_p k_B U}
\right]\\
& -\frac{\left(T_{\alpha}-T_p\right)}{T_p}
\left[
\frac{\nu_{\alpha p}}{U}
+\frac{\nu_{p \alpha}}{U}\frac{T_\alpha}{T_p} \right].
\label{eqn:dEpsilon.1}
\end{split}
\end{align}
To model the excess temperature ratio beyond $R_b$, we assume that
either $Q_s = 0$ for both ion species, or that any remaining heating
affects both species equally. This allows to relate $Q_p$ and
$Q_\alpha$ by
\begin{equation}
Q_\alpha= Q_p \frac{n_\alpha T_\alpha}{n_p T_p}
\label{eqn:Qs}
\end{equation}
which upon insertion into Eqn~\ref{eqn:dEpsilon.1} allows us to neglect
the input heating terms. We further assume that both ion species
follow the same radial density profile
\begin{equation}
n_s(r) \propto n_{0s}r^{-\xi}
\label{eqn:density}
\end{equation}
leading to the cancellation of the $dn_s/dr$ terms in
Eqn.~\ref{eqn:dEpsilon.1}. With these two assumptions, we have the
simplified expression
\begin{align}
\begin{split}
\frac{d \epsilon}{d r} = &
-\epsilon \left[
\frac{\nu_{\alpha p}}{U}
+\frac{\nu_{p \alpha}}{U} \left(\epsilon +1\right)
\right] =-\frac{\nu_{\alpha p}}{U}\left[
\epsilon \left(1 + F\right)
+ \epsilon^2 F
\right]
\label{eqn:dEpsilon.2}
\end{split}
\end{align}
where we have employed standard expressions for the Coulomb collision
frequency between two Maxwellian distributions, 
\begin{equation}
\nu_{ss'}=4 \pi q_s^2 q_{s'}^2 \frac{{\rm ln} \Lambda n_{s'}}{m_s \mu w_{ss'}^3}
\label{eqn:nu}
\end{equation}
presented in \cite{hernandez87} with reduced mass ratio $\mu \equiv
m_sm_{s'}/(m_s + m_{s'})$ and the combined thermal speed $w_{ss'}^2 = 2
T_s/m_s + 2 T_{s'}/m_{s'}$, to write the ratio of collision
frequencies in terms of the mass density ratio
\begin{equation}
\frac{\nu_{p \alpha}}{\nu_{\alpha p}}=
\frac{n_\alpha m_\alpha}{n_p m_p} \equiv F.
\label{eqn:F}
\end{equation}
As \cite{Tracy:2015} recently demonstrated, for all heavy ions in the
solar wind including He$^{2+}$ the dominant coupling via Coulomb
collisions is with H$^+$.

As $\nu_{\alpha p}$ depends on both $T_\alpha$ and $T_p$, separating
$\epsilon$ and $\nu_{\alpha p}$ as necessary for a solution to
Eqn.~\ref{eqn:dEpsilon.2} necessitates the construction of a 'reduced'
collision frequency which only depends on a single temperature
\begin{align}
\tilde{\nu}_{ss'}= & 8 \pi q_s^2 q_{s'}^2 \frac{{\rm ln} \Lambda
  n_{s'}}{m_s^2 w_{s'}^3}
=\frac{2 {\nu}_{ss'}}{1+m_s/m_{s'}}
\left(1+\frac{T_sm_{s'}}{T_{s'}m_s}\right)^{3/2}
\label{eqn:nu0}
\end{align}
where the single species thermal speed is $w_{s'}^2 = 2
T_{s'}/m_{s'}$. Using this reduced collision frequency, we separate
Eqn.~\ref{eqn:dEpsilon.2} into terms which do and do not depend on
$\epsilon$, resulting in a differential equation of the form
\begin{equation}
\int_{R_b}^{R_w}\frac{2}{5} \frac{\left [1 + \frac{\left( \epsilon + 1
      \right)}{4} \right]^{3/2}d \epsilon} {\epsilon \left (1 + F
  \right) + \epsilon^2 F} =
-\int_{R_b}^{R_w}\frac{\tilde{\nu}_{\alpha p}(r)}{U(r)} dr \equiv -A_c
\label{eqn:deq}
\end{equation}
where our solution depends on integration from the outer boundary of
the zone of preferential heating $R_b$ to the radius of the observer
$R_w$.

Expanding the numerator and performing typical $u$-substitutions,
known integral identities, and arithmetic manipulations yields a
closed form expression for the left-hand side of Eqn.~\ref{eqn:deq}:
\begin{widetext}
\begin{align}
\begin{split}
\int_{R_b}^{R_w} \frac{2}{5} \frac{\left [1 + \frac{\left( \epsilon + 1 \right)}{4} \right]^{3/2}d \epsilon}
{\epsilon \left (1 + F \right) + \epsilon^2 F}
= 
&\frac{1}{10}\frac{\sqrt{5 + \epsilon_w}-\sqrt{5 + \epsilon_0} }{F} -\frac{\sqrt{5}}{2\left(1+F\right)}\left[ 
\frac{1}{2} {\rm \ln} \left(\frac{\sqrt{1+\frac{\epsilon_w}{5}} + 1}{\sqrt{1+\frac{\epsilon_w}{5}} - 1}\right)
-
\frac{1}{2} {\rm \ln} \left(\frac{\sqrt{1+\frac{\epsilon_0}{5}} + 1}{\sqrt{1+\frac{\epsilon_0}{5}} - 1}\right)
\right]\\
&+\frac{\left(4F -1\right)^{3/2}}{10F^{3/2}\left(1+F\right)}
\left[\arctanh\left(\frac{\sqrt{F}\sqrt{5+\epsilon_w}}{\sqrt{4F-1}}\right)
-\arctanh\left(\frac{\sqrt{F}\sqrt{5+\epsilon_0}}{\sqrt{4F-1}}\right)\right]
\label{eqn:solution}
\end{split}
\end{align}
\end{widetext}
where $\epsilon_0$ and $\epsilon_w$ are the excess temperature ratio at the outer boundary of the zone of preferential heating and at 1 AU respectively.

A solution for $A_c$, the right-hand side of Eqn.~\ref{eqn:deq},
requires a description for the radial evolution of the reduced
collision frequency $\tilde{\nu}_{\alpha p}$ which depends on the radial
structure of $n_p$, $U$, and $T_p$, as well as a value for the
boundary $R_b$. From Eqn.~\ref{eqn:nu0}, $\tilde{\nu}_{\alpha p}$
varies as $n_pT_p^{-3/2}$.  Both $n_p$ and $T_p$ fall off with
distance from the Sun, so it is expected that $\tilde{\nu}_{\alpha,
  p}$ should increase substantially closer to the Sun.  Using the
radial variations found in Helios observations \citep{Hellinger:2013},
we take $T_p\propto r^{-\delta}$, $U\propto r^{-\sigma}$and
$n_p\propto r^{-2}U(r)$.  From these scalings, $\tilde{\nu}_{\alpha
  p}$ as a function of the measured collision rate at 1 AU
$\tilde{\nu}_{\alpha p}^w=\tilde{\nu}_{\alpha p}(R_w)$ may be written
as
\begin{equation}
\tilde{\nu}_{\alpha p}= \tilde{\nu}_{\alpha p}^w \left(\frac{R_w}{r}
\right)^{2+\sigma -1.5 \delta}
\label{eqn:ratio}
\end{equation}
and thus the collisional age integral is expressed as
\begin{align}
\begin{split}
A_c = & \int_{R_b}^{R_w} dr \frac{\tilde{\nu}_{\alpha p}^w}{U_w} \left(
\frac{R_w}{r}\right)^{2+2\sigma - 1.5 \delta}\\ = &
\frac{-\tilde{\nu}_{\alpha p}^w}{1+2\sigma - 1.5 \delta}\frac{R_w}{U_w}
\left[1-\left(\frac{R_b}{R_w}\right)^{-1-2\sigma + 1.5 \delta}
  \right].
\label{eqn:Ac}
\end{split}
\end{align}
Note that care must be taken in evaluating this equation, as a
singularity appears for $-2\sigma + 1.5 \delta=1$.  We note the
assumed scaling relations used for $\tilde{\nu}_{\alpha p}$ may not be
accurate especially close to the Sun as temperatures in the corona are
lower than extrapolations from the {Helios} trend lines close to
the Sun.  We will offer a {post hoc} justification that $R_b$ is
sufficiently far from the Sun where these scaling relations serve as
accurate descriptions.

Combining Eqns.~\ref{eqn:solution} and \ref{eqn:Ac} into
Eqn.~\ref{eqn:deq}, one can produce a transcendental expression that
relates $\epsilon_w$ to measured quantities $F, n_p,U,T_p$, fixed
parameters $\delta, \sigma$, and free parameters $R_b,
\epsilon_0$. Note that rather than determining the parameters that
result in $\epsilon_w$ approach zero for high $A_c$, we allow our solution to relax to a residual
$\epsilon_1$, which is treated as a free parameter in our modeling, to
account for the fact that $\epsilon\simeq 0.2-0.3$ has been reported
even in the case of high $A_c$ for both {Wind}/SWE observations of
helium and hydrogen temperatures \citep{Maruca:2013a,Kasper:2013} and
for heavier ions \citep{Tracy:2016}. It is an open question whether
this residual $\epsilon_1$ is indicative of continuing preferential
heating acting in interplanetary space at a much reduced weaker level
compared to in the zone of preferential heating, or if it is indicative
of an instrumental measurement error in temperature ratios.  The
asymptotic value of the temperature excess below $R_b$ in the zone will therefore
be $\epsilon_0 + \epsilon_1$.  This modeled value $\epsilon_w$ will be
compared to observed values of $\epsilon$ in the fashion described in
the following section in an effort to indirectly measure the extent of
the zone of preferential heating.

\section{Determination of Zonal Boundary using the {Wind} Data Set}
\label{sec:Rb}
We now describe our procedure for solving for the outer boundary $R_b$
of the zone of preferential heating by comparing our model predictions
for the excess temperature ratio with observations of the solar wind
by the {Wind} spacecraft.  The model is a function of solar wind
speed, density, temperature, mass density ratio $F$, and spacecraft
location for each measurement, along with five global free parameters:
the boundary height $R_b$, the steady state excess temperature ratio
within the zone $\epsilon_0+\epsilon_1$, the residual excess at 1 AU
$\epsilon_1$, and the radial exponents of solar wind speed $\sigma$
and temperature $\delta$.  We use observations of the radial
dependence of solar wind properties from the {Helios} mission to guide
our choice of exponents, since as we will show, our values for $R_b$
are closer to {Helios} perihelion than they are coronal heights where
there are spectroscopic measurements.  Since there are different
reported values for the radial temperature exponent $\delta$ in the
literature \citep{marsch82,Hellinger:2013}, we consider four values of
$\delta$, $0.7, 0.8, 0.9, 1.0$ that cover the reported range.
 Those same studies have also shown that $\delta$ may be a
  weak function of speed, so the observations were analyzed in
  separate $25$ km s$^{-1}$ intervals in solar wind speed.  While
  Fig.~\ref{fig:tap_vp_ac} clearly shows the same relaxation of
  $\epsilon$ with increasing Coulomb collisions continues at least to
  $650 $ km s$^{-1}$, we have limited our analysis to the range
  $300-425$ km s$^{-1}$ in order to have good observational statistics
  at high and low $A_c$ with $25$ km s$^{-1}$ interval size.   For this
  work, we keep the solar wind speed exponent fixed at
  $\sigma=0$. Observational studies have found negligible dependencies
  of the solar wind speed on radial distance, with a preference toward
  a shallow increase in $U$ with $r$
  \citep{Hellinger:2011,Hellinger:2013}. As the speed and temperature
  exponents only appear as a linear combination in the expression for
  $A_c$, exploring the dependence of $R_b$ on $\delta$ also provides
  direct insight into the effects of changing $\sigma$.

For each range in solar wind speed and assumed value of $\delta$ and
$\sigma$ we now determine the best fit values of our three free
parameters $\epsilon_0$, $\epsilon_1$, and $R_b$ by minimizing the
$\chi^2$ per degree of freedom difference between the model and the
observations, weighted by an error estimate.  It might seem like the
easiest way to conduct this analysis would be to directly compare the
predicted and observed $\epsilon$ for every individual observation in
a given speed interval.  We found that this was unreliable, as
Fig.~\ref{fig:tap_vp_ac} shows there is a very strong preference for a
particular collisional age at a given speed.  In order to avoid
biasing the analysis due to the most common age, we instead histogram
our observations into bins in $A_c$, calculate the mean $\epsilon$ in
each bin, and compare those means to the model value.

We start with an initial guess for $R_b$, $\epsilon_0$, and
$\epsilon_1$.  We found that the following analysis is highly
insensitive to those initial guesses, but fitting a simple exponential
curve to the data provided a good initial guess that speeds up the
calculations.  For each measurement of $\epsilon_w$ we then calculate
an initial collisional age $A_c^i$ using the current values for $R_b$,
$\epsilon_0$, and $\epsilon_1$. We also calculate an overall average
mass density ratio $F$ using the selected data. The measured
$\epsilon$ are then binned as a function of the calculated $A_c^i$,
with the range of $A_c$ and the resolution of our bins set up
beforehand so there are always at least $1,000$ individual
measurements per bin. We use the average over all the selected data.
We then calculate a prediction for $\epsilon_w$ for each bin using the
transcendental expression resulting from Eqns.~\ref{eqn:deq},
\ref{eqn:solution} and \ref{eqn:Ac}. We do not want to use the
uncertainty in the mean for each interval in $A_c$ for this analysis
because the high correlation between speed and $A_c$ shown in Fig. 2
would strongly bias the best fit to the handful of intervals with the
bulk of the observations.  Instead we identified a constant error
estimate based on the observed spread of $\epsilon$ at very high
$A_c$.  We found that at high $A_c$ the majority of epsilon
observations are normally distributed with a width of 7\%, and used
this value as the error estimate at all $A_c$.  The non-linear best
fit is thus calculated by variation of $R_b$, $\epsilon_0$, and
$\epsilon_1$ and iteration of the above binning routine, producing an
estimate for the global minimum of $\chi^2/dof$ along with the best
fit values and one-sigma uncertainties for the three free parameters
for each interval of $U$ and value of $\delta$.

\begin{figure}
  \includegraphics[width=2.25in,angle=90]{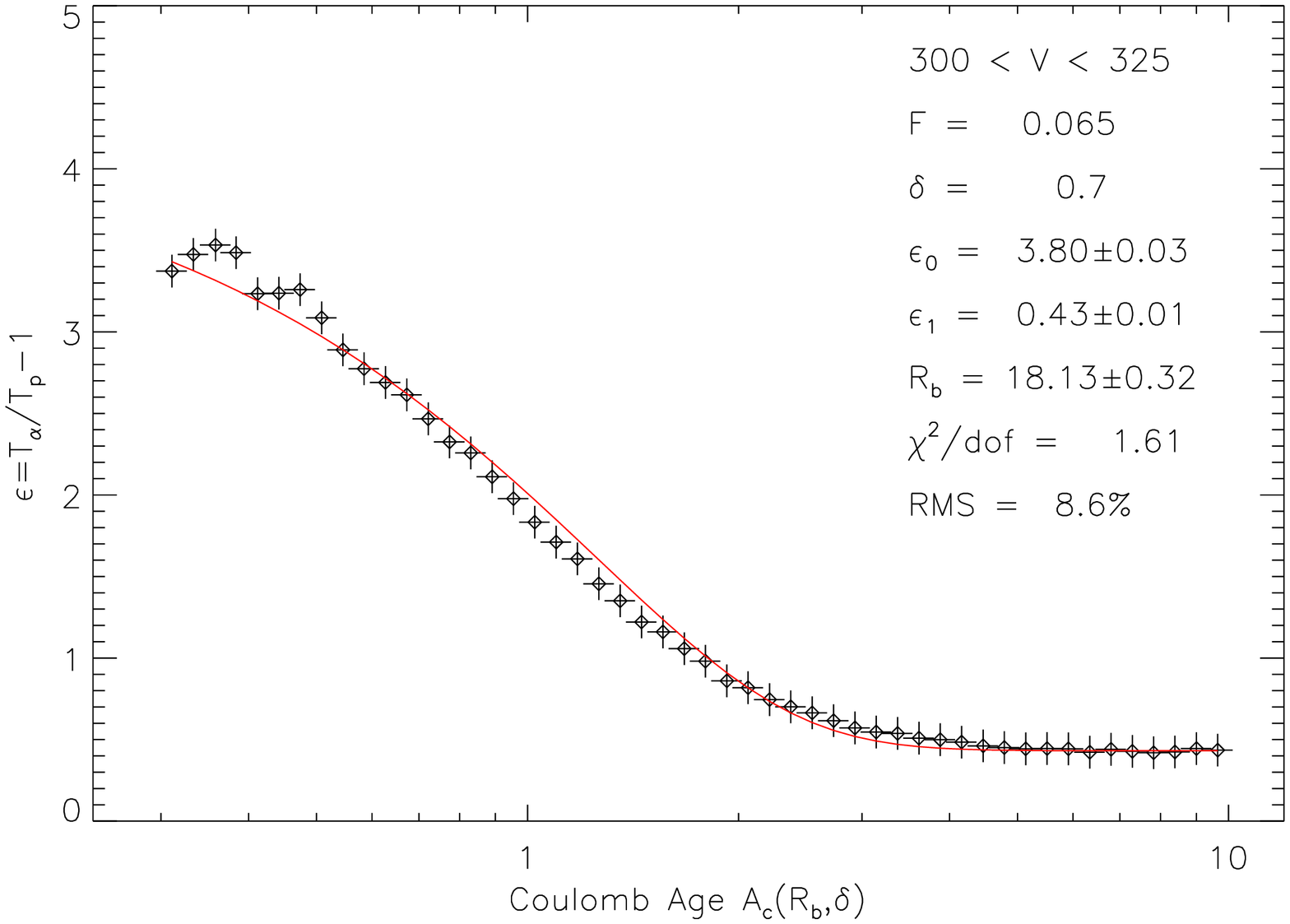}\\
  \includegraphics[width=2.25in,angle=90]{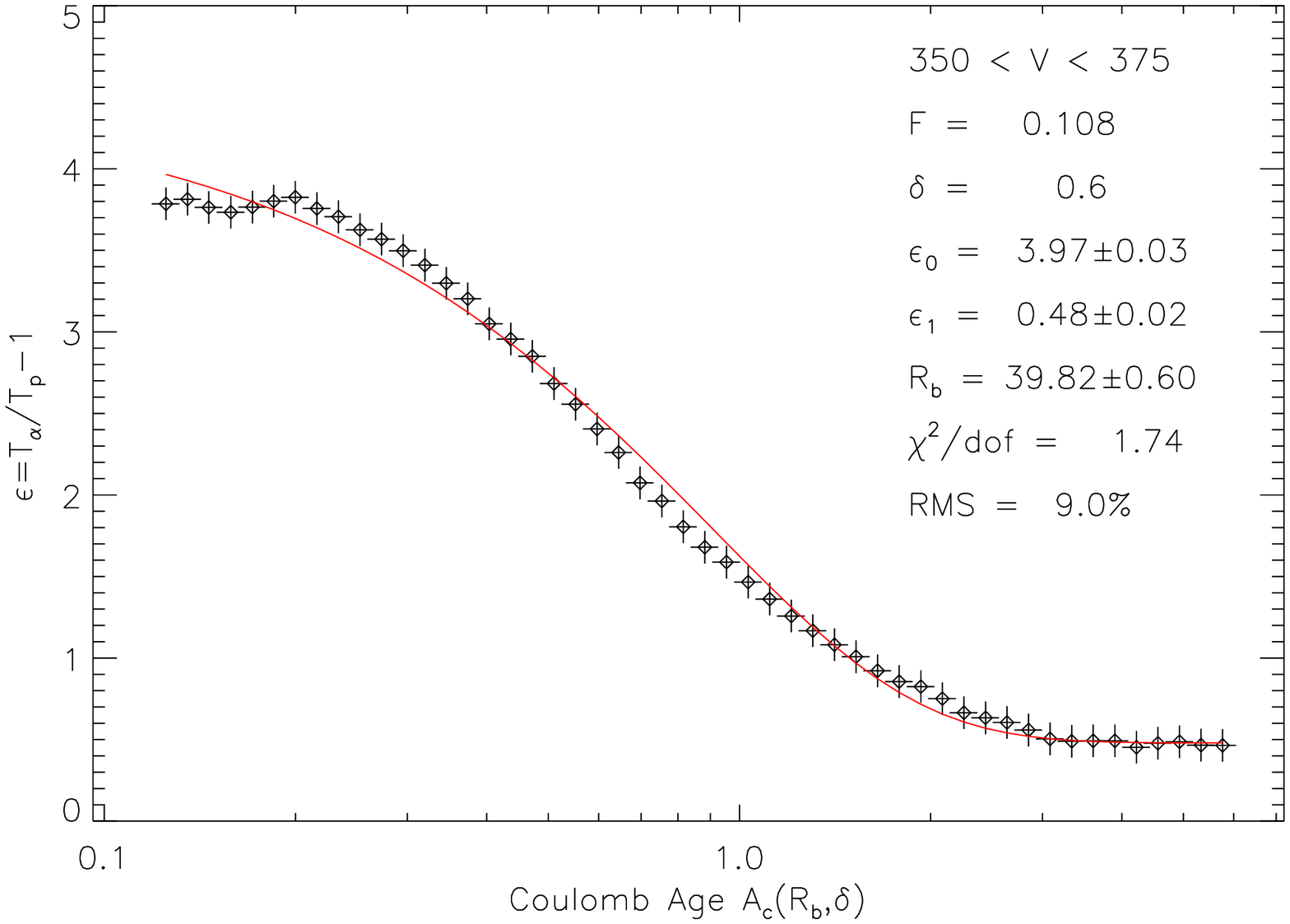}\\
  \includegraphics[width=2.25in,angle=90]{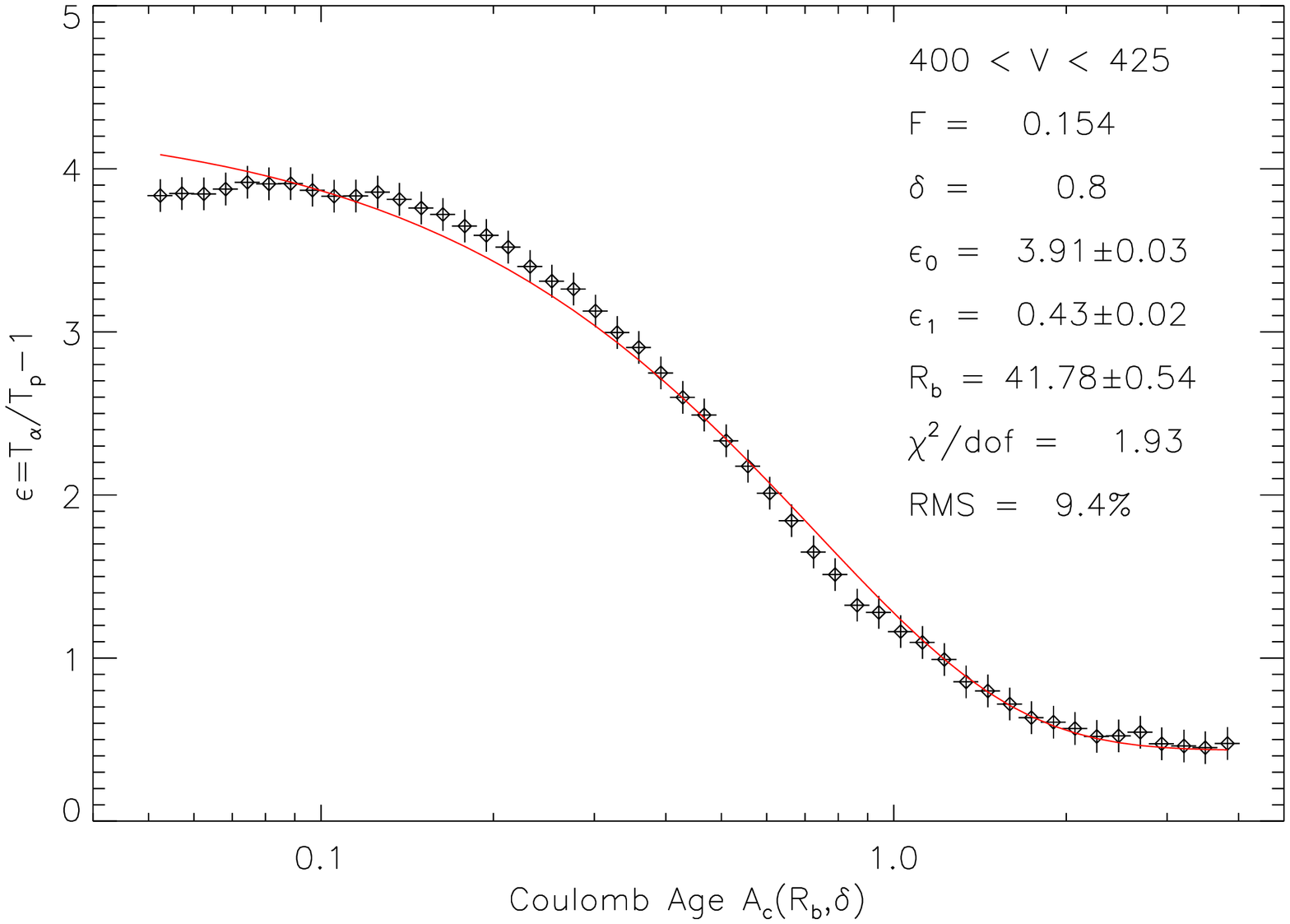}\\
  \caption{Best fit match (red line) of Eqn.~\ref{eqn:deq} to solar
    wind observations (diamonds) of helium temperature excess relative
    to hydrogen for three intervals in solar wind speed. Legends
    indicate the mean mass density ratio, assumed radial temperature
    exponent $\delta$, best fit values for the zone boundary $R_b$,
    excess temperature ratios $\epsilon_0$ and $\epsilon_1$. Also
    shown are $\chi^2/dof$ and the RMS deviation between the model and
    the observations. }
\label{fig:findrq}
\end{figure}

Fig. \ref{fig:findrq} illustrates the process and results for three
solar wind speed intervals $U=300-325, 350-375, 400-425$ km s$^{-1}$,
using $\delta=0.7,0.6,0.8$ respectively. Our simple model of a zone of
preferential heating is able to predict the mean excess helium
temperature to within 8-9\% with a $\chi^2/dof=1.6-1.9$. Typical
values for $R_b$ are tens of solar radii from the Sun.

\section{Discussion} \label{sec:discuss}

\begin{figure}
\includegraphics[width=2.5in,angle=90]{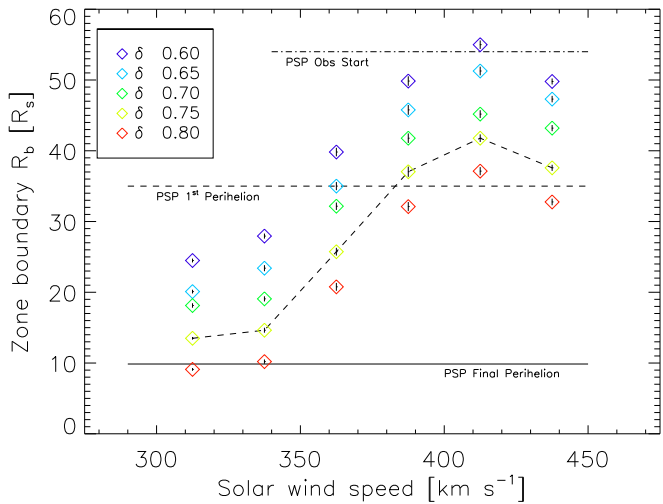}
  \caption{Diamonds, with error bars, indicate the best fit values for
    the outer boundary $R_b$ of the zone of strong preferential
    heating close to the Sun as a function of solar wind speed and for
    four different values for the radial temperature power law
    exponent $\delta$.  Using $\delta=0.75$ from Helios observations
    in the inner heliosphere, solar wind with speeds from 300-425 km
    s$^{-1}$ experience strong non-thermal heating to an outer
    boundary $10-35$ $R_s$ from the Sun (dashed path).  Horizontal
    lines indicate the start of science observations (dot-dash line),
    the first perihelion distance at the start of the mission
    (dashed), and the closest perihelion distance at the end of the
    mission for {Parker Solar Probe}.  Based on these results, we
    predict that \PSP will be the first spacecraft to enter and
    directly observe this zone of preferential non-thermal heating.}
\label{fig:rq_speed}
\end{figure}

\begin{figure}
\includegraphics[width=2.5in, angle=90]{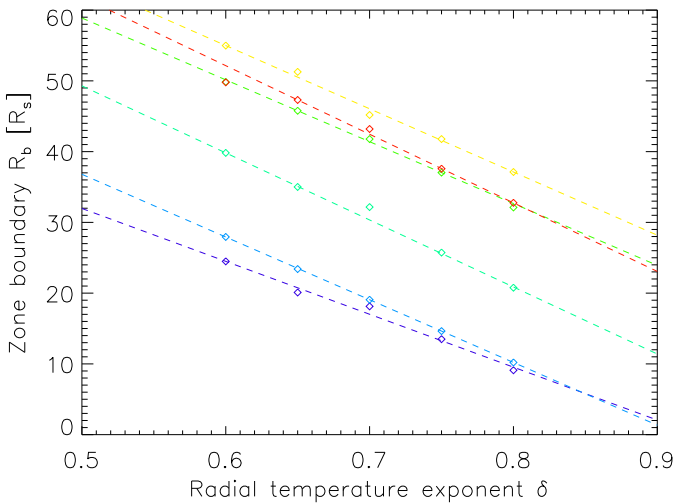}
  \caption{Scaling of best $R_b$ with $\delta$ for different solar
    wind speeds.  Color from purple to red indicates increasing solar
    wind speed.  In general there is a simple linear dependence
    between $\delta$ and $R_b$, with the best fit extent of the zone
    dropping 8.8 $R_s$ for an $0.1$ increase in $\delta$.}
\label{fig:epsilon_delta}
\end{figure}

Following the procedure outlined in the previous section, the best fit
values for $R_b$ as a function of solar wind speed and temperature
power law exponent are calculated and shown in
Fig.~\ref{fig:rq_speed}. For the value of $\delta=0.75$, matching the
observations of slow wind reported in \citet{Hellinger:2013}, the
outer edge of the boundary ranges from $15$ to $40 \ R_{\rm S}$ from
the Sun's surface for varying solar wind speed. The zone boundary is a
decreasing function of $\delta$, and as shown in
Fig.~\ref{fig:epsilon_delta} we see that there is a simple linear
relationship that allows one to correct $R_b$ for different
assumptions of $\delta$, with the dependence of the boundary value on
$\delta$ fairly independent of speed.  Averaging over all the trends
shown in Fig.~\ref{fig:epsilon_delta}, $R_b$ drops by $8.8 \ R_{\rm
  s}$ for every $0.1$ increase in $\delta$. Physically, the faster
temperature falls off with distance from the Sun, the smaller the
preferential heating zone. Similarly, a linear relation between $U$
and $R_b$ can be found approximately satisfying $R_b \propto 0.1 U$,
not shown.

Perhaps the most significant implication of the inferred zone is that
the preferential heating does not persist throughout the heliosphere,
and thus can not be measured locally by spacecraft at 1 AU. Attempts
to locally differentiate between proposed mechanisms for the
preferential heating will rely on missions such as \PSP
\citep{Fox:2015} and \SO \citep{Muller:2013} that will make
measurements of particles and fields in the near-Sun region of the
heliosphere \citep{Kasper:2015,Bale:2016}. Several key radial
distances for the \PSP mission are shown in Fig.~\ref{fig:rq_speed}.
All but one of the predicted $R_b$ are below the starting
distance for \PSP science observations.  By the end of the mission
\PSP will cross all but one of the predicted values of
$R_b$, allowing direct measurement of the region where the
preferential heating is predicted to occur.

We note that our assumption that $R_b$ is sufficiently far away to
employ radial scalings of $T_p$ and $U$ measured by {Helios} has
been justified {post hoc}. Had $R_b$ been on the order of a few
solar radii, a more sophisticated model for the radial dependence of
the collision frequency would have become necessary. Additional
modifications to this model, such as the inclusion of the effects of
temperature anisotropy or other non-thermal features
\citep{Hellinger:2016} on the collision frequency may have a
quantitative effect on the position of $R_b$, but we expect these
effects to be small. Additionally, we will be able to improve our
model by using measurements of the radial dependence of $n$, $U$, and
$T$ by \PSP to determine the accuracy of power-law extrapolations from
{Helios} observations into the near-Sun environment.

When evaluating this model, one must address the nature of the energy
input beyond $R_b$.  While the structure of this model and the
best-fit values of $R_b$ indicate that the preferential heating is
limited to a region close to the Sun, this does not necessarily imply
that no heating persists beyond this region. We know that ion heating
of some level extends out to 1 AU and beyond \citep{Cranmer:2009}, but
the rate most likely drops with distance.  In the radial model of
\citet{Chandran:2011} for example, the heating rate is high until
about 20 $R_s$ and then it falls off as a power law. Our model allows
for such heating as long as the heat input per particle for the
$\alpha$ particles satisfies Eqn.~\ref{eqn:Qs}.  The persistence of a
residual, non-zero $\epsilon$ even for high-$A_c$ plasma, commented on
in \cite{Maruca:2013b}, may be an indication of a small amount of
preferential heating of ions beyond $R_b$, or may be an instrumental
limitation. Characterization of this residual excess temperature ratio
will be left to future work.

We think the most plausible interpretation of our results
  is that $R_b$ is linked to the \Alfven critical point $R_A$, the
  radial distance where the solar wind transitions from being
  sub-\Alfvenic to super-Alfv\'enic, and that the zone of preferential
  heating is simply the volume of space below the \Alfven point where
  reflected waves traveling back towards the Sun interact with
  escaping waves to enhance the turbulent cascade and allow it to
  transport significant energy in the form of intense and
  counter-propagating fluctuations down to ion kinetic scales.
  Typical predicted values for $R_A$ lie between $10$ and $30$ $R_s$
  \citep{Verdini:2012,Perez:2013}, consistent with our findings for
  $R_b$.  Since all sunward directed \Alfvenic fluctuations generated
  below $R_A$ are trapped below $R_A$, it is natural to expect that
  intense reflection driven turbulence will be stronger in this region
  \citet{Verdini:2007}. 

As a specific example of a preferential ion heating mechanism that
would be active below $R_A$ in the presence of counter-propagating ion
kinetic scale fluctuations consider \citet{Kasper:2013}, which showed
that when $A_c$ is small the dependence of $T_\alpha/T_p$ on plasma
$\beta$ and normalized differential flow speed $\Delta V_{\alpha
  p}/C_A$ is consistent with heating by counter-propagating Alfven
ion-cyclotron waves (AIC, kinetic scale Alfvenic fluctuations
propagating in opposite directions along the local magnetic field)
which are significantly more efficient at heating $He^{2+}$ relative
to $H^+$ and in fact all other ions heavier than $H^+$.  In the
presence of a spectrum of AIC waves propagating in a single direction,
ions heavier than $H^+$ are slowly heated as resonant scattering
diffuses them in velocity space about the phase speed of the waves.
If counter propagating waves are introduced, $He^{2+}$ and heavier
ions can scatter off waves traveling in opposite directions,
permitting a more general diffusion in velocity space and a far more
rapid and preferential heating.  A unified explanation of the
observations could be as follows.  Everywhere below the \Alfven point,
counter-propagating \Alfven waves are present and strongly
preferentially heat ions heavier than $H^+$.  This heating is first
apparent $0.1-0.3 \ R_S$ above the surface of the Sun when the Coulomb
collisions are no longer able to prevent temperature differences from
emerging.  From that height up to the \Alfven point all ions are
heated by these counter-propagating waves and diffuse in phase space
to reach an equilibrium temperature excess.  Suddenly at the \Alfven
point reflected waves are not able to travel back towards the Sun, and
the power in counter-propagating waves drops significantly, shutting
off the counter-propagating AIC mechanism.  The temperature excess
developed below the \Alfven point then decays through Coulomb
relaxation to the level observed at an interplanetary spacecraft.
This sharp drop in heating at $R_A$ is consistent with our assumption
that heating stops at $R_b$ and could help explain why our simple
model for heating with distance fits the observations so well.  In
this framework the reason \citet{Kasper:2013} could only see their
correlations with AIC predictions in low $A_c$ solar wind is because
they were never directly observing the heating in action, but instead
a signature frozen into solar wind ions that crossed $R_A$, and
therefore $R_b$, days earlier.  Finally, the small residual temperature
excess seen for ions in high $A_c$ plasma \citep{Kasper:2013,
  Maruca:2012, Tracy:2016} could then be due to the weaker heating of
ions by AIC fluctuations traveling predominantly in one direction.

Another intriguing possibility is that $R_b$ could
  correspond to the distance recently identified in
  \citet{DeForest:2016} where a transition from relatively steady and
  laminar radial flow to sheared and turbulently mixed flow is
  remotely observed.  Using an improved analysis of remote
  observations from the Heliospheric Imager on {STEREO}, the authors
  identified a region $\sim 40-80 \ R_s$ from the solar surface in
  which the smooth radial expansion of the slow solar wind appears to
  fragment and break up.  As with the \Alfven critical point, the
  transition in solar wind flow structure reported by
  \citet{DeForest:2016} could signify another boundary in the solar
  wind that separates different levels of fluctuations and dominant
  heating mechanisms. Of course it is also possible that the breakup
  in smooth flow seen in the images is simply a manifestation of the
  \Alfven point itself, or another height where the \Alfven mach
  number crosses some value and the plasma becomes unstable.

\begin{figure}
\includegraphics[width=2.5in, angle=90]{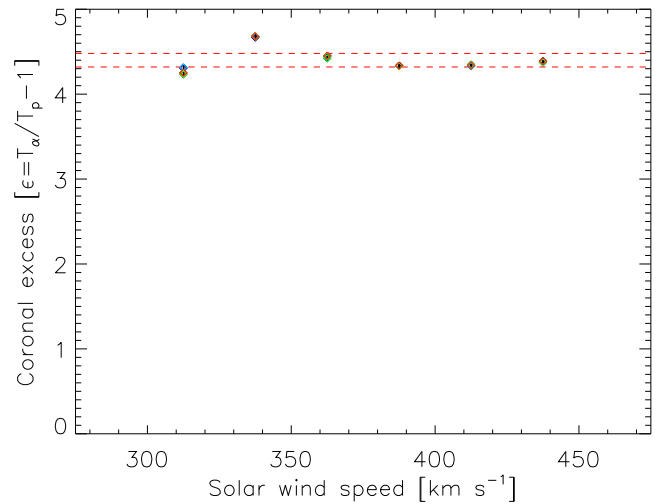}
  \caption{Best fit values for the excess helium temperature relative
    to hydrogen $\epsilon_0+\epsilon_1$ in the zone of preferential
    heating as a function of solar wind speed.  Different color
    symbols indicate assumed value for $\delta$ and are
    indistinguishable, indicating that there is a single steady state
    temperature ratio for ions within the preferential heating zone
    independent of speed.  Dashed red horizontal lines indicate
    predicted helium excess based on fits to the mass dependence of
    heavy ions in collisionless solar wind by \cite{Tracy:2016},
    showing the combined studies are consistent with there being a
    single super-mass-proportional scaling of $T_i/T_p=(4/3)m_i/m_p$
    resulting from preferential heating acting within the zone.}
\label{fig:epsilon_excess}
\end{figure}

Lastly, we can use the values for $\epsilon_0+\epsilon_1$ to look at
the implied excess temperature ratio back in the zone of preferential
ion heating for direct comparison with coronal heating theories and
with other observations. Fig.~\ref{fig:epsilon_excess} shows the
coronal excess temperature of helium relative to hydrogen as a
function of solar wind speed and temperature exponent $\delta$, with
the symbols for $R_b$ at different $\delta$ using the same color
scheme as Fig.~\ref{fig:epsilon_delta}.  Here we see another
significant signature of the physical process responsible for
preferential ion heating in the inner heliosphere.  Our analysis shows
an average temperature excess of about $4.2-4.3$, meaning that that
helium is $5.2-5.3$ times hotter than hydrogen, independent of solar
wind speed and our assumption for $\delta$.  This excess would appear
to be a highly significant and model independent result.  Any theory
of heating in the corona or extended solar wind should be able to
produce this steady state excess temperature.  Finally, we note the
dashed red lines on the figure, which were developed by taking the
temperature dependency in low $A_c$ or collisionless solar wind
reported by \cite{Tracy:2016} for heavy ions in the solar wind, but
evaluated for the mass of helium. We find that within the error
reported in that analysis, the implied steady state helium temperature
excess in the corona is consistent with the mass dependence of heavy
ions in the solar wind when selecting collisionless wind that
presumably indicates the coronal values.  We therefore propose that our
results, combined with those of \cite{Tracy:2016}, are consistent with
the idea that there is a preferential ion heating mechanism acting in
a zone of non-thermal heating, that acts on all ions, and extends out
tens of solar radii from the Sun. Within this zone ion temperatures
reach a steady state ratio of $T_i/T_p=(4/3) m_i/m_p$, independent of
solar wind speed.

\section{Conclusion} \label{sec:cncl}

We have examined the temperature ratio of fully ionized He$^{2+}$ and
H$^+$ in the solar wind and its dependence on Coulomb collisional age
in order to solve for the location of an outer boundary of an apparent
zone of preferential ion heating in the inner heliosphere. Using
millions of observations from the {Wind} spacecraft in concert
with a physically motivated model for the excess temperature ratio, we
are able to construct a best fit value for the outer boundary of this
region, which falls between $\sim 20 - 40 \ R_S$ with some variation
with solar wind speed and radial temperature scalings.  The restricted
radial extent of this region would frustrate attempts to identify
preferential heating mechanisms using measurements at 1 AU, but future
missions including \PSP will provide measurements both within and
outside this region, allowing for the novel measurement of the
mechanisms that lead to the non-thermal heating of solar wind minor
ions.

We can now answer the three questions proposed in the
  Introduction.  The large unequal ion temperatures seen in situ by
  spacecraft in the solar wind are not maintained by ongoing local and
  strong preferential heating.  Instead they are a leftover of heating
  that happened closer to the Sun.  Solar wind at all speeds appear to
  experience strong preferential heating within our proposed zone, and
  the only reason fast wind appears more non-thermal than slow wind in
  interplanetary space is due to the large difference in Coulomb
  collisions that transpire as the solar wind travels from the outer
  boundary of the zone to the observing spacecraft.  The strong
  preferential ion heating seen close to the Sun in spectroscopic
  observations continues $\sim 20 - 40 \ R_S$ from the Sun before
  dropping off, perhaps due to a lack of counter-propagating Alfvenic
  fluctuations.  It is possible that the residual temperature excess
  observed in interplanetary space indicates that a weaker form of
  preferential heating is active outside of the zone, but it is only
  able to produce temperatures that are different by tens of percent.

\section*{Acknowledgments} The authors acknowledge support from the {Wind} team
for the data used for this project.  J.C. Kasper was supported by NASA
Grant NNX14AR78G. K.G. Klein was supported by NASA HSR Grant
NNX16AM23G.

\vspace*{.5cm}





\end{document}